\documentclass[aps,prd,onecolumn,groupedaddress,showpacs,nofootinbib,amssymb]{revtex4}
\usepackage[dvips]{graphicx}
\usepackage{amssymb}
\usepackage{amsmath}
\usepackage{graphicx}
\usepackage{amsfonts}
\usepackage{bm}
\begin{document}

\title{Type IV singular bouncing cosmology from $f(T)$ Gravity}
\author{
P.~H.~Logbo,$^{(a,b)}$\,\footnote{Email: pascoloo@yahoo.fr}
M.~J.~S.~Houndjo,$^{(a,b)}$\,\footnote{Email: sthoundjo@yahoo.fr}}
\affiliation{ $^{a}$\, Facult\'e des Sciences et Techniques de Natitingou, BP 72, Natitingou, B\'enin \\
$^b$ \, Institut de Math\'{e}matiques et de Sciences Physiques, 01 BP 613,  Porto-Novo, B\'{e}nin
}
\begin{abstract}
We investigate bouncing scenario in the modified $f(T)$ gravity, $T$ being the torsion scalar. Attention is attached to the reconstruction of $f(T)$ able to describe type IV singular bouncing evolution, where we adopt as assumption that the bouncing and type IV singularity points coincide. In the context of the reconstructed $f(T)$ model we calculate the Hubble slow-roll parameters in order to determine the dynamical evolution of the cosmological system under study. The results show that the Hubble slow-roll parameters become singular at the type IV singularity indicating a dynamical instability. Moreover we perform the stability analysis of the $f(T)$ gravity solution where, according to the obtained result, the type IV singularity point is a saddle point in agreement with the bounce scenario. 
\end{abstract}

\pacs{04.50.Kd, 95.36.+x, 98.80.-k, 98.80.Cq}

\maketitle


\def\pp{{\, \mid \hskip -1.5mm =}}
\def\cL{\mathcal{L}}
\def\be{\begin{equation}}
\def\ee{\end{equation}}
\def\bea{\begin{eqnarray}}
\def\eea{\end{eqnarray}}
\def\tr{\mathrm{tr}\, }
\def\nn{\nonumber \\}
\def\e{\mathrm{e}}

\section{Introduction}

Bouncing cosmology \cite{1deOi,2deOi,3deOi,4deOi,5deOi,6deOi,7deOi,8deOi,9deOi,10deOi} is an important way to solve initial singularity problem, which in the view of cosmological theories, is an undesirable characteristic.  According to the bouncing cosmology the universe presents a contracted phase which ends up to a minimal radius, and after that starts an expanded phase. This means that the collapsing feature of the universe never occurs, reveling the impossibility of initial singularity. In general, the cosmological bounce feature can be observed in two main cases, the first case being the Loop Quantum Cosmology \cite{10deOi} matter bounce theories \cite{12deOi} and the second in scalar theories where singularities should appear \cite{1deOi,2deOi,3deOi,4deOi,5deOi,6deOi,7deOi,8deOi,9deOi,12deOi}.  On the other hand modified bouncing cosmology feature can be explained making use of modified theories of gravity \cite{10deOi,13deOi}. \par
As we can see in the literature, initial singularity is one of many other types of cosmological singularities, the so-called finite time singularities \cite{20deOi,21deOi,22deOi,23deOi,24deOi,25deOi,26deOi}. In general Hawking-Penrose theorems describe consistently all type of singularities \cite{14deOik}.  The spacelike singularities in general are a direct consequence of gravitational collapse of massive objects \cite{15deOik},  while  timelike singularities have a more cosmic effect. In theories based on curvature it is usually noted  that spacetime singularities are crushing types where the curvature scalar strongly diverges. Among the finite time singularity the Big Rip appears as he most severe due to its violent phenomenological feature \cite{17deOik}, while for the types II, III and type IV singularities the situation is not the same. The type I, II and III singularities are characterized by  divergent physical quantities at the singular point, while the type IV singularity does not produce catastrophic feature with regard to the observable physical quantities.  Type IV singularity has been the subject of several studies, see \cite{16deOik, 18deOik,19deOik,20deOik,21deOik,22deOik,23deOik,24deOik,25deOik}, where among other interesting results, it appears that closed universe satisfying the strong energy conditions $\rho +3p$ and $\rho>0$, need not to recollapse \cite{Oik,26deOik,27deOik}. It is also shown in \cite{22deOik,23deOik,24deOik} that type IV singularity, just affects the dynamical evolution of the cosmological system, but does not have effect  of the physical observable quantities. Referring to the affected dynamics, we mean the slow-roll parameters which determine the dynamical evolution of the cosmological system \cite{28deOik}. On the other way, the slow-roll parameters can possibly diverge leading to a dynamically stable system \cite{21deOik,22deOik,23deOik}. \par 
In this paper, we attach attention to cosmological bounce model including type IV  finite time singularity that can be generated from a general pure $f(T)$ gravity. We assume that the bouncing point is the same where the type IV singularity occurs, and find out the algebraic $f(T)$ function accordingly. Note that this kind of work has been developed in $f(R)$ gravity \cite{Oi,Oik} and \cite{14deOi,37deOi,38deOi,39deOi,40deOi,41deOi}, where interesting results have been obtained. More precisely, in \cite{Oi} authors investigated which Jordan frame $f(R)$ gravity can describe a type IV singular bouncing cosmological evolution. In our present paper, the attempt is almost the same but the work is developed in the framework of $f(T)$ theory of gravity. The important assumption we shall make is that the bouncing point coincides with the point where the  type IV singularity occurs. In particular, we will find the $f(T)$model which generates the bounce feature near the finite time type IV singularity point.  After that, attention will be attach to the study of stability of the obtained model.\par
The paper is organized as follow: the Sec. \ref{sec2} is devoted to the classification of the finite time singularities, while Sec. \ref{sec3} focuses on  the description of bounce solution. The bounce solution in the context of $f(T)$ gravity is developed in Sec. \ref{sec4}
and the related Hubble slow-roll parameters for type IV singular evolution are performed in Sec. \ref{sec5}. 
The stability analysis of the reconstructed $f(T)$ solution near the bounce is realized in Sec. \ref{sec6} and the conclusion in Sec. \ref{sec7}.

\section{Finite-time singularities}\label{sec2}
Finite-time cosmological singularities have been formally classified in Refs. \cite{25deOi,26deOi}. There are essentially four types of finite time, known as Type I, II, III and Type IV singularities, classified as follows \cite{25deOi, 26deOi}.\par
$\bullet$ Type I (Known as Big Rip Singularity): This type of cosmological singularity is the most severe among
finite time cosmological singularities. It occurs when, as the cosmic time approaches specific time $(t_s)$, the effective energy density $\rho_{eff}$, the scale factor $a(t)$, and
also the effective pressure $p_{eff}$ diverge. \par
$\bullet$ Type II (Known as Sudden Singularity): This singularity occurs when, as the cosmic time approaches
 $t_s$, only the scale factor $a(t)$ and the effective energy density $\rho_{eff}$ take bounded values, while the effective pressure diverges. This means that the second and higher derivatives of the scale factor diverge.\par
$\bullet$ Type III: This singularity occurs when, as the cosmic time approaches $t_s$, only the scale factor remains
finite, but both the effective energy density and the corresponding effective pressure diverge. \par
$\bullet$ Type IV: This type of singularity is the most mild among all the three aforementioned types of finite time
singularities, and we shall focus on this type of singularity in the following. This singularity occurs when, as the cosmic time approaches $t_s$, all the cosmological physical quantities remain finite,  but the higher derivatives of the Hubble rate diverge.\par
We are interested to the Type IV singularity due to its particular feature referring to the geodesic incompleteness issue that occurs at finite time singularities. That is not the case for Type I and initial singularity, where their occurrence are very severe, violating the energy conditions. Note that the type IV singularity is less-harmful because there is no occurrence of geodesic incompleteness, showing that this singularity is not so serious. However their occurrence and possible consequences should be examined in order to fully understand this type of singularity and what it can yield. This paper is devoted to the study of bouncing cosmology including in its evolution a type IV singularity. The situation is such that no initial singularity can appear and the only that able to occur during the bounce evolution is the type IV. Motivated by these properties of the type IV singularity, we propose to point out possible realization of non-singular bouncing cosmology in the framework of $f(T)$ gravity.
\section{Describing bounce solution} \label{sec3}
The cosmological bounce is based on the succession of two phases; a contraction era and an expansion era. In the contracting phase, the scale factor decreases such that $\dot a< 0$, and this until the universe reaches a minimal radius where $\dot a =0$; leading the transition, after what the expansion phase starts i.e, the scale factor increases ($\dot a>0$). This phenomenon of transiting from a contracting phase to an expanding one render the appealing bouncing cosmology \cite{7deOi,9deOi,12deOi}, then, avoiding the standard description with inflationary models.\par
Assuming that the bouncing point occurs at a cosmic time $t_s$, the contracting phase corresponds to $t<t_s$ and the Hubble parameter $H(t)<0$, while for the expanding phase, $t>t_s$ with $H(t)>0$. As in the paper \cite{Oi}, we shall consider the scale factor characterizing the singular bounce as
\begin{eqnarray}\label{scalefactor}
a(t)=e^{f_0(t-t_s)^{2(1+\epsilon)}},
\end{eqnarray}
where $\epsilon$ and $f_0$ are specific constants. 
This expression of the scale factor is chosen such that it is normalized to one at the bounce. The corresponding Hubble rate reads
\begin{eqnarray}
H(t)=2f_0(1+\epsilon)\left(t-t_s\right)^{2\epsilon+1}\label{Hrate}\;.
\end{eqnarray}
According to the classification of finite time singularities, for the type IV singularity the exponent in Eq. (\ref{scalefactor}) has to be greater than $1$, i.e, $2\epsilon+1>1$, such that $\epsilon>0$. Let us set $\beta=2\epsilon+1$ for simplicity in such away that 
\begin{eqnarray}
H(t)=2f_0(1+\epsilon)\left(t-t_s\right)^{\beta}\,.\label{hubblerate}
\end{eqnarray}
Then, it appears that, the type I, II, III and IV singularities occur for $\beta<-1$, $0<\beta<1$, $-1<\beta<0$ and $\beta>1$, respectively. \par 
Working in Planck unit system, the Hubble rate is measured in $eV$, the cosmic time in $(eV)^{-1}$, such that the parameter $f_0$ is measured in $(eV)^{2\epsilon+2}$. On the other, cosmic time shall be expressed in second $sec$, the Hubble rate in $(sec)^{-1}$ and $f_0$ in $(sec)^{-2\epsilon-2}$. These units will be adopted in the rest of the paper. From the classification of the singularities, it is obvious that using (\ref{hubblerate}), Type IV singularity occurs for $2\epsilon +1>1$, i.e, for all positive values of $\epsilon$. 
 As assumed in \cite{Oi}, on other hand,  one shall assume that $\epsilon<1$ in which the expression (\ref{scalefactor}) appears as a small deformation of the well known bouncing cosmology \cite{14deOi}
 \begin{eqnarray}
 a(t)=e^{f_0(t-t_s)^2}.\label{scalefactor2}
 \end{eqnarray}
 It is clear that the type IV singularity occurs at $t=t_s$, knowing as the bouncing point. \par 
 From (\ref{Hrate}), it appears that the type IV singularity occurs when the higher derivative of the Hubble rate diverges, i.e,
 \begin{eqnarray}
 \frac{d^nH(t)}{dt^n}\rightarrow\infty,
 \end{eqnarray}
 for $n\geq 2$. For $n=2$ the Hubble rate diverges for $1<\beta<2$. When $\beta>2$ the Hubble rate becomes finite and the divergence is observed on the third derivative, i.e, for $n=3$. It is clear that the singularity occurs for $0<\epsilon<\frac{1}{2}$, guaranteeing that the bounce (\ref{scalefactor}) is a deformation of (\ref{scalefactor2}).\par

\section{Bounce solution in $f(T)$ gravity} \label{sec4}
Our task in this section is to find the algebraic $f(T)$ gravity able to generate cosmological evolution of singular bounce with scale factor (\ref{scalefactor}). We also assume total absence of matter fluids, and will make use of the standard reconstruction scheme \cite{39deOi}-\cite{41deOi}, giving special attention to what happens near the type IV singularity. Let us start with the general pure $f(T)$ action
\begin{eqnarray}\label{action}
S=\int d^4x\,e\,f(T)
\end{eqnarray}
After varying the action (\ref{action}) with respect to the tetrad, and making use of the flat Friedmann-Robertson-Walker metric, the equation of continuity reads
\begin{eqnarray}
-36H^2\dot{H} f_{TT}+3\left(4H^2+\dot{H}\right)f_T+f=0\,.
\end{eqnarray}
We introduce an auxiliary scalar field $\phi$ in order to perform the reconstruction scheme, rewriting the action as
\begin{eqnarray}\label{actionphi}
S=\int d^4x \;e \left[P(\phi)T+Q(\phi)\right]\,.
\end{eqnarray}
We assume absence of kinetic term of $\phi$ such that, by varying (\ref{actionphi}) with respect to $\phi$, one gets
\begin{eqnarray}\label{varyingresult}
P'(\phi)T+Q'(\phi)=0,
\end{eqnarray}
where $P'(\phi)$ and $Q'(\phi)$ denote the derivatives of $P$ and $Q$ with respect to $\phi$, respectively. The resolution of (\ref{varyingresult}), when it is possible, leads to $\phi(T)$, such that the reconstructed algebraic will be presented  as
\begin{eqnarray}\label{fphiT}
f(\phi(T))=P(\phi(T))T+Q(\phi(T)).
\end{eqnarray}
We now need to find the expressions $P(\phi(T))$ and $Q(\phi(T))$, and to do so, we written down the FRW equations of field as
\begin{eqnarray}
12H^2P(\phi)+P(\phi)T+Q(\phi)&=&0\label{eq1}\\
-48H^2\dot{H}P'(\phi)+4(3H^2+\dot{H})P(\phi)+P(\phi)T+Q(\phi)&=&0\label{eq2}\;.
\end{eqnarray}
By eliminating $Q(\phi)$ from (\ref{eq1})-(\ref{eq2}) and setting the auxiliary scalar field to the cosmic time $t$, one gets
\begin{eqnarray}\label{eqdiff}
12H^2\dot{P}-P=0\,,
\end{eqnarray}
whose solution reads
\begin{eqnarray}\label{solP}
P(t)=C_1\exp{\left[\alpha_1(t-ts)^{1-2\beta}\right]}\;,\quad\quad \alpha_1=\frac{1}{48f_0^2(1-2\beta)(1+\epsilon)^2}\;,
\end{eqnarray}
where $C_1$ in an integration constant. By injecting (\ref{solP}) into (\ref{eq1}), one obtains
\begin{eqnarray}\label{solQ}
Q(t)=-24C_1f_0^2(1+\epsilon)^2(t-t_s)^{2\beta}\exp{\left[\alpha_1(t-ts)^{1-2\beta}\right]}
\end{eqnarray}
Remembering that the torsion scalar is related to the Hubble parameter through $T=-6H^2$, and making use of (\ref{fphiT}), (\ref{solP})-(\ref{solQ}), one gets
\begin{eqnarray}
f(T)=2C_1T\exp{\left[ \alpha\left(- T \right)^{\frac{1-2\beta}{2\beta}} \right]}\;;\quad\quad \alpha=\alpha_1\left[24f_0^2(1+\epsilon)^2\right]^{\frac{2\beta -1}{2\beta}}\;.
\end{eqnarray}
\section{The Hubble Slow-Roll Parameters for Type IV Singular Evolution} \label{sec5}
In this section we propose to calculate the slow-roll parameters usually used in the cosmology and called  Hubble slow-roll parameters \cite{Oik,28deOik}. In the context of the $f(T)$ reconstructed model we are interested to find the behavior of the slow-roll parameters near the type IV singularity. These parameters determine the dynamical evolution of the cosmological system and denote them as $\varepsilon_H$ and $\eta_H$ \cite{28deOik},
\begin{eqnarray}
\varepsilon_H=-\frac{\dot{H}}{H^2}\;,\quad\eta_H=-\frac{\ddot{H}}{2H\dot{H}}\;.
\end{eqnarray}
Making use of (\ref{hubblerate}), one gets
\begin{eqnarray}
\varepsilon_H= -\frac{\beta(t-t_s)^{-\beta -1}}{2f_0(1+\epsilon)}\label{hubbleparat1}
\end{eqnarray}
and 
\begin{eqnarray}
\eta_H= -\frac{(\beta-1)(t-t_s)^{-\beta -1}}{4f_0(1+\epsilon)} \label{hubbleparat2}
\end{eqnarray}
An elementary analysis of (\ref{hubbleparat1}) and (\ref{hubbleparat2}) shows that as the cosmic time $t$ goes toward  the type IV singularity time $t_s$ the Hubble slow-roll parameters diverge (because $\beta>1$  $\Rightarrow$ $-\beta -1<-2$ ), leading to a strong instability. On the other hand it can be noted that type IV singularity does not lead to infinite physical quantities at the the singularity time is approached, in contrast with other type type of singularities. However, the effect of type IV singularity is expressed at the level of the dynamical evolution
and abruptly stopped at the singularity point.\par

\section{Analyzing the stability of reconstructed $f(T)$ model}\label{sec6}
\begin{figure}[h]
	\centering
	\begin{tabular}{rl}
		\includegraphics[width=8cm, height=8cm]{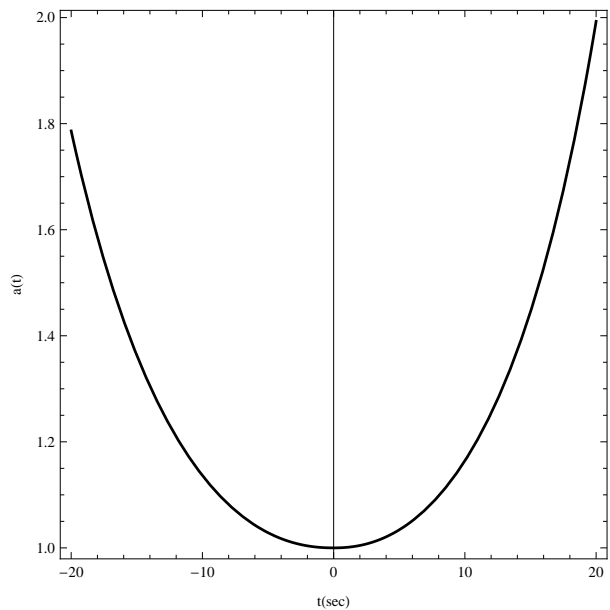}
	\end{tabular}
	\caption{ This figure shows the evolution of the scale factor $a(t)$ as the cosmic time $t$ evolves, for $t_s=10^{-35}$ sec, $\epsilon=\frac{1}{11}$ and $f_0=10^{-4} (sec)^{-2\epsilon-2}$, with scale factor (\ref{scalefactor}). }
	\label{fig1}
\end{figure}

\begin{figure}[h]
	\centering
	\begin{tabular}{rl}
		\includegraphics[width=8cm, height=8cm]{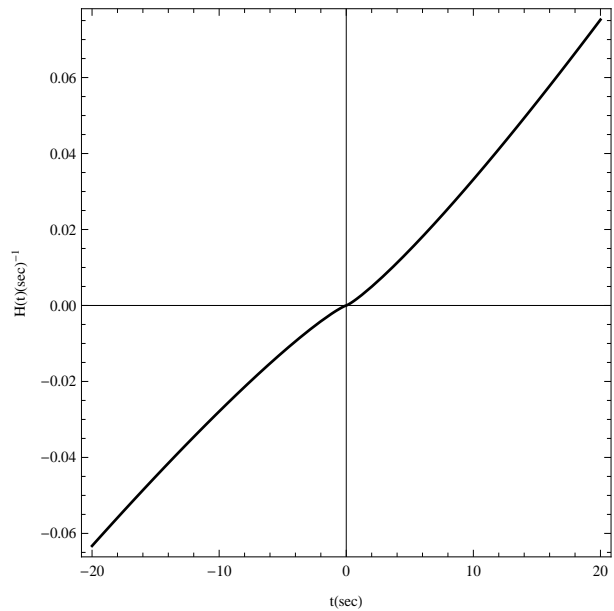}
	\end{tabular}
	\caption{ Evolution of the Hubble parameter $H(t)(sec)^{-1}$ as the cosmic time $t$ evolves, for $t_s=10^{-50}$ sec, $\epsilon=\frac{1}{11}$ and $f_0=10^{-4} (sec)^{-2\epsilon-2}$, using (\ref{Hrate}).  }
	\label{fig2}
\end{figure}

\begin{figure}[h]
	\centering
	\begin{tabular}{rl}
		\includegraphics[width=8cm, height=8cm]{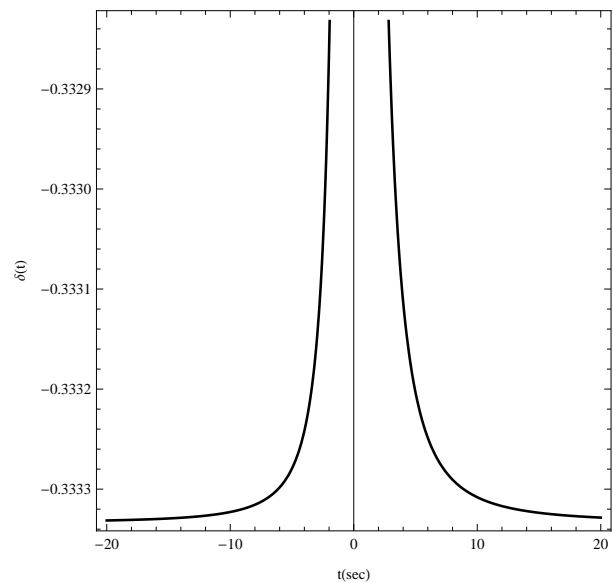}
	\end{tabular}
	\caption{ Evolution of the perturbation function $\delta(t)$ as the cosmic time $t$ evolves, for $t_s=10^{-50}$ sec, $\epsilon=\frac{1}{11}$.}
	\label{fig3}
\end{figure}

Let us now investigate the stability of the obtained solution. To do so, let us assume the scale factor as \cite{39deOi}-\cite{41deOi}
\begin{eqnarray}
a=a_0e^{g(\phi)}\;,
\end{eqnarray}
where $g(\phi)=\left(\phi-t_s\right)^{2(\epsilon+1)}$. Thus (\ref{eqdiff}) takes the following form
\begin{eqnarray}
12(g'(\phi))^2\frac{dP(\phi)}{d\phi}\left(\frac{d\phi}{dt}\right)^3-P(\phi)=0\;.\label{eqdiff2}
\end{eqnarray}
We introduce the function $\delta$  defined as
\begin{eqnarray}
\delta = \frac{d\phi}{dt} - 1\,,\label{pertdef}
\end{eqnarray}
view as the parameter that measures the exact way that perturbations behave. By injecting (\ref{pertdef}) in (\ref{eqdiff2}), one gets
\begin{eqnarray}
\delta(t)=\frac{P(\phi)}{36\left[g'(\phi)\right]^2\frac{dP}{d\phi}}-\frac{1}{3}|_{\phi=t}\;,
\end{eqnarray}
which, in term of auxiliary field $\phi$ and cosmic time, takes the following respective forms
\begin{eqnarray}
\delta(\phi)=\frac{(\phi-t_s)^{-2(1+2\epsilon)}}{144(1+\epsilon)^2}-\frac{1}{3}\\
\delta(t)=\frac{(t-t_s)^{-2(1+2\epsilon)}}{144(1+\epsilon)^2}-\frac{1}{3}
\end{eqnarray}
Before proceeding let us address another point of view where one may observe a possible inconsistency. As we it has been shown in the previous section, we assume $0<\epsilon<1/2$, and then setting \cite{Oi}
\begin{eqnarray}
(t-t_s)^{2(\epsilon+1)}=(t-t_s)^{\frac{2n}{2m+1}}\,,
\end{eqnarray}
where the arbitrary parameters $m$ and $n$ have to appropriately be chosen so that $0<\epsilon<1/2$. As developed in \cite{Oi}, fixing $m=5$ and $n=12$ leads to a non-complex scale factor, and we write it as
\begin{eqnarray}
a(t)=e^{f_0(t-t_s)^{\frac{24}{11}}} _,,
\end{eqnarray}
from which we extract $\epsilon = \frac{1}{11}$. Thus, the Hubble rate takes the following form
\begin{eqnarray}
H(t)=\frac{24}{11}f_0(t-t_s)^{\frac{22}{11}}(t-t_s)
\end{eqnarray}
We now propose to observe the behavior the both the scale factor and the Hubble rate as the  cosmic time evolves. To do so, we fix the singularity time $t_s=10^{-50} sec$, $\epsilon=\frac{1}{11}$ and $f_0=10^{-4}(sec)^{-2\epsilon-2}$. The scale factor and the Hubble rate are illustrated by Fig.\,\ref{fig1} and Fig.\,\ref{fig2}, respectively. The scale factor is decreasing (contracting) before the bounce, reaches a minimal value at the bounce ($t=t_s$), and is increasing (expanding) after the bounce. Concerning the Hubble rate, it is negative before the bounce, vanishing at the bounce point and becomes positive after. We then see that all qualitative features of the bounce are satisfied. For more analysis, we plot the perturbation function $\delta(t)$, illustrated by Fig.\,\ref{fig3}, and see that before the bounce it increases, meaning that the dynamical system is unstable. After the bounce the perturbation function decreases going toward almost vanishing values, synonym of stable dynamical system. The simplicity of the analysis of the perturbation function in this paper, in contrast with the one done in \cite{Oi} based on the $f(R)$, is favored by the fact that field equation in $f(T)$ is more manipulable.  



\newpage

\section{Conclusion}\label{sec7}
This paper is devoted to the study of bounce cosmology with type IV singularity appearing  at the bouncing point in the context of $f(T)$ theory of gravity. We attach attention to cosmic times around  the type IV singularity by reconstructing the $f(T)$ able to describe the bounce near this singularity.  As well known, the type IV singularity is not a violent finite time singularity, unlike the type I or the initial singularity, such that all physical quantities defined on spacelike hypersurface at the bounce are finite and the effects of this singularity cannot be observed at the level of physical observable quantities. Moreover we analyze the Hubble  slow-roll parameters in order to determine the dynamical evolution of the cosmological system. Our results show that near the singularity the Hubble slow-roll parameters diverge, and it appears here that the effects of this singularity can be governed by these parameters. \par
Having in hand the reconstructed $f(T)$ gravity near the bounce, we also analyze the stability of the solution near the bounce. We see that when the universe is contracting, the perturbation function increases  leading to an unstable dynamical evolution of the cosmological system, while for the expanding phase the perturbation function decreases and goes to toward infinitesimal small values, meaning that the cosmological system is now in a stable dynamical phase. The point common to the bounce and type IV singularity is then a saddle point.



\end{document}